\documentclass[]{aastex631}

\shorttitle{quasi-periodic micro-jets}
\shortauthors{Hong et al.}
\graphicspath{{./}{figures/}}
\begin{document}

\title{Quasi-periodic microjets driven by granular advection as observed with \\ high-resolution imaging at He \text{\small{I}} 10830 \AA}

\author{Zhenxiang Hong}
\affiliation{Key Laboratory of Dark Matter and Space Astronomy, Purple Mountain Observatory, CAS, Nanjing, 210023, China, wangya@pmo.ac.cn}
\affiliation{School of Astronomy and Space Science, University of Science and Technology of China, Hefei 230026, China}
\author{Ya Wang}
\affiliation{Key Laboratory of Dark Matter and Space Astronomy, Purple Mountain Observatory, CAS, Nanjing, 210023, China, wangya@pmo.ac.cn}
\author{Haisheng Ji}
\affiliation{Key Laboratory of Dark Matter and Space Astronomy, Purple Mountain Observatory, CAS, Nanjing, 210023, China, wangya@pmo.ac.cn}

\begin{abstract}
With high-resolution narrowband He \text{\small{I}} 10830 \AA\ filtergrams from GST,  we give an extensive analysis for 4 granular sized  microeruptions which appear as the gentle ejection of material in He \text{\small{I}} 10830 \AA\ band.  The analysis was aided with the EUV data from AIA and line-of-sight magnetograms from HMI on board SDO. The microeruptions are situated on magnetic polarity inversion lines (PILs), and their roots are accurately traced down to intergranular lanes. Their durations are different, two microeruptions are repetitive microjets, lasting $\sim$ 50 and 27 minutes respectively, while the other two events are singular, lasting $\sim$ 5 minutes. For the two microjets, they are continuous and recurrent in  He \text{\small{I}} 10830 \AA\ band, and the recurrence is quasi-periodic with the period of  $\sim$ 5 minutes. We found that only transient co-spatial EUV brightenings are observed for the longer duration microjets and EUV brightenings are absent for the two singular microeruptions. What is essential to the longer duration microjets is that granules with the concentration of positive magnetic field persistently transport the magnetic field to the PILs, canceling the opposite magnetic flux and making the base of the two microjets and the underlying granules migrate with the speed of $\sim$ 0.25 and 1.0 km s$^{-1}$. The observations support the scenario of  magnetic reconnection for the quasi-periodic microjets and further show that the reconnection continuously generates multi-temperature components, especially the cool component with chromospheric temperature. In addition, the ongoing reconnection is modulated by p-mode oscillations inside the Sun.
\end{abstract}


\section{Introduction} \label{sec:intro}

Even in solar quiet regions, the upper atmosphere is a highly dynamic environment, full of microeruptions, as seen from high resolution observations. A typical phenomenon is the upward ejection of material, presumably being ejected along magnetic field lines.  They were first observed as surges with H$\alpha$ telescopes \citep{Roy1973b}.  Surges are events of plasma ejection from the chromosphere into the corona, and they were found to be associated with EUV and soft X-ray emission \citep{Schmieder1988, Schmahl1981, Svestka1990,Shen2021}. With ongoing improvement in spatial and temporal resolutions, and especially observations at multiple wavelengths, we have observed various facets of the upward material ejection in the solar upper atmosphere, and they were named as X-ray jet \citep{Shibata1992}, EUV jet \citep{Chae1999}, and H$\alpha$ jet \citep{Chae1998, Wang1998ha} from the viewing point of a certain wavelength. \\

In terms of their locations, jets can be classified as polar jets \citep{Wang1998pj,Culhane2007,Sterling2015,Panesar2018,McGlasson2019}, active region jets \citep{Chifor2008} and quiet region jets \citep{Panesar2016,McGlasson2019,Panesar2020}.  Some larger EUV/X-ray jets, especially those polar jets, can escape into space and appear in the white-light coronagraph, and they sometimes are termed as coronal jets  \citep{StCyr1997,Paraschiv2015}. Recent observations show that a great number of coronal jets are made by minifilament eruptions that are prepared and triggered by  flux cancellation at a magnetic neutral line, especially for those in quiet regions and coronal holes \citep{Panesar2016,Panesar2018,Panesar2020,McGlasson2019}.  For most solar eruptions that make a coronal jet, an essential component is a long spire that extends out along far-reaching unipolar magnetic field rooted in an anemone-shaped base. The base depicts a closed-field magnetic
configuration over a patch of opposite-polarity magnetic flux. The spire is evidently made by external reconnection of a lobe of the jet's magnetic-anemone base with the ambient oppositely-directed far-reaching field. On the other hand,  spicules are one class of jet phenomena on the solar limb traditionally observed in narrowband H$\alpha$ images \citep{Beckers1968}, and they may appear as dark mottles, fibrils (including dynamic fibrils) on the disk \citep{Tsiropoula1994, Suematsu1995, Christopoulou2001}.  \cite{DePontieu2007} found that there are two fundamentally different types of spicules at the limb,  type I and  type II spicules. The type I spicules evolve on timescales of several minutes while type II spicules occur on timescales of a few tens of seconds during which they are seen to rise and then rapidly disappear. Type I spicules seem to correspond to dynamic fibrils (active regions) and a subset of mottles (quiet Sun), when seen on solar disk \citep{Hansteen2006, DePontieu2007b, Rouppe2007}. Short-lived type II spicules may correspond to rapid blue-shifted events on the disk \citep{Langangen2008}.  Spicules are also thought to play an important role in heating the upper atmosphere \citep{Samanta2019} and many small-scale jets are sources of mass and energy of solar wind \citep{Tian2014}.\\

Jet activities, either hot or cool, tend to be recurrent from a small localized region \citep{Chae1999, Jiang2007, Chifor2008, Guo2013, Chen2015, Chandra2015}.  Some authors even proposed periodic nature for them.  \cite{Jiang2007} reported three surges with a recurrent period of about 70 minutes in H$\alpha$, EUV, and soft X-ray. With wavelet analysis, \cite{Li2015} obtained the periods of 5 and 13 minutes from intensity variations at the base of a series of recurrent EUV jets. With high-resolution narrowband imaging at He \text{\small{I}} 10830 \AA, \cite{Wang2021} reported periodic behavior of $\sim$ 5 minutes period for a small-scale jet. However, in the EUV band and for the same jet, the periodic behavior vanishes and appears as non-periodic recurrent brightenings. In this regard,  high-resolution narrowband imaging at He \text{\small{I}} 10830 \AA\ is a relatively new window for various jets in the chromosphere. \\ 

With the 1.6-meter aperture Goode Solar Telescope at Big Bear Solar Observatory (BBSO/NST), a new facet of solar upper chromosphere was obtained using high-resolution narrowband imaging at He \text{\small{I}} 10830 \AA. Meanwhile, the first direct observations of ultra-fine jet-like dynamical events originating in solar inter-granular lanes and subsequently lighting up the corona were observed \citep{Ji2012}. With the same set of data, further interesting results have been achieved. For completeness of this paper, they are listed as follows: 
\begin{itemize}
\item \cite{Zeng2013} reported the association of a He \text{\small{I}} 10830 \AA\ surge with the magnetic reconnection driven by advection of a rapidly developing large granule. 
\item In a quiet region, \cite{Hong2017} found that EUV emission is mainly concentrated in inter-granular lane areas where He \text{\small{I}} 10830 \AA\ absorption is enhanced. 
\item In a EUV moss region, numerous small-scale material ejections shown as enhanced He \text{\small{I}} 10830 \AA\ absorption are squirted from inter-granular lanes periodically with the period of $\sim$ 5 minutes.  Furthermore, periodic material ejection was found to be associated with periodic EUV brightenings and magnetoacoustic oscillations with roughly the same period \citep{Ji2021, Hashim2021}. 
\end{itemize}
Therefore, we need to know how the scale size spectrum of coronal jets extends down to these small-scale sub-arcsecond material ejections when high-resolution observations are available \citep{Raouafi2016, Hou2021}.  We also need to know the behaviors of all above-mentioned microeruptions in high-resolution He \text{\small{I}} 10830 \AA\ imaging observations.  In this paper, we extensively analyzed 4 microeruptions  that occurred at 4 different places with the same set of data. The sizes of these He \text{\small{I}} 10830 \AA\ microeruptions are comparable with granules' sizes. With simultaneous and well-aligned high-resolution images of the photosphere, we are able to pin down these microeruptions in the context of advecting granules and corresponding magnetic field.  \\

\section{Observations} \label{sec:obs}

On  July 22, 2011, the active region NOAA 11259 was observed with narrowband (bandpass: 0.5 \AA) imaging at the blue wing  (-0.25 \AA) of He \text{\small{I}}  10830 \AA\ and broadband imaging (bandpass: 10 \AA) centered at TiO 7057 \AA\ line using Goode Solar Telescope (GST) and its facilities at Big Bear Solar Observatory (BBSO) \citep{Cao2010, Goode2010}. It is the same set of data analyzed by \cite{Ji2012}. The images of He \text{\small{I}} 10830 \AA\ have a pixel size of $\sim$ $0.09{^\prime}{^\prime}$ and cadence of $\sim$ 14 seconds. For TiO 7057 \AA\ images, they have a pixel size of $\sim$ $0.036{^\prime}{^\prime}$ and a cadence of $\sim$ 15 seconds. The EUV/UV imaging data taken by the Atmospheric Imaging Assembly (AIA; \citealt{Lemen2012}) and line of sight (LOS) magnetograms taken by the Helioseismic and Magnetic Imager (HMI; \citealt{Scherrer2012}) on board Solar Dynamics Observatory (SDO) are used to show the hot components of these microjets and relevant variations of magnetic field.  \\

The smallest events analyzed in this paper have a footpoint size of $\sim$ $0.5{^\prime}{^\prime}$. To understand their physical originations, we have to pin down their locations on the high-resolution photospheric images in the context of granules and inter-granular lanes, a precise co-alignment between the images of the photosphere and chromosphere is very necessary. He \text{\small{I}} 10830 \AA\ is a shallow chromospheric line, which allows the photosphere to be visible at most places of the high-resolution narrowband images as a background. This attribute makes us to align the images with the images of TiO 7057 \AA\ much easier.  Meanwhile, the co-alignment between GST and SDO data was carried out using sunspots, pores, and bright granules on TiO 7057 \AA\ broadband images and HMI continuum images as the intermediary features.  Remaining offset after the co-alignment is less than $\sim$ $0.5{^\prime}{^\prime}$ \citep{Hong2017}.  \\

\section{Results} \label{sec:res}

\begin{figure}
\epsscale{1}
\plotone{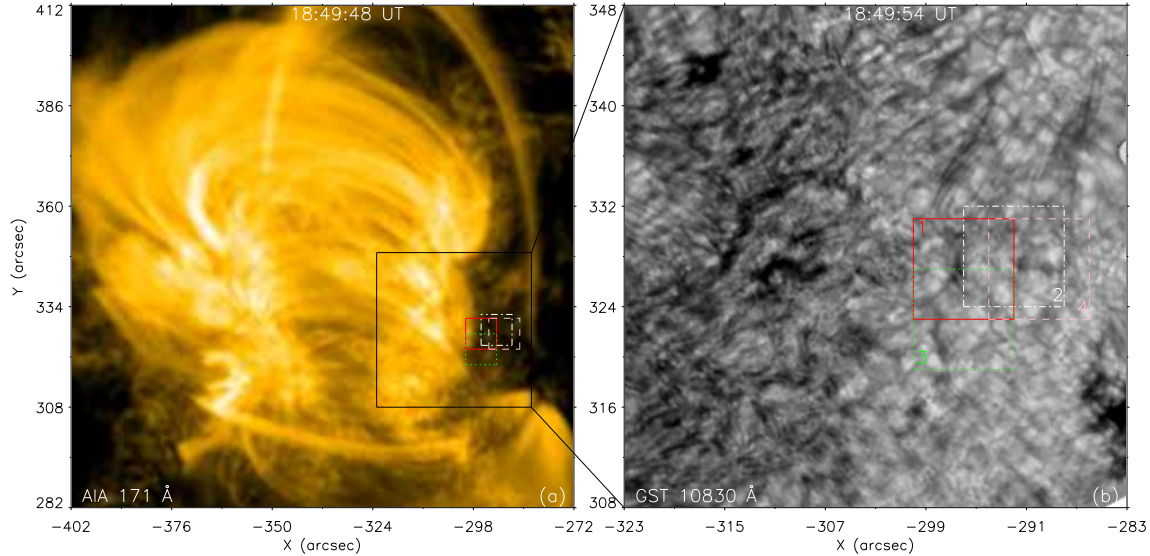}
\caption{Panel (a) depicts an overview of the active region NOAA 11259 as observed by AIA at 171 \AA, while panel (b) is a sample He \text{\small{I}} 10830 \AA\ narrowband filtergram which FOV is given in panel (a) with a black box.  Four smaller boxes with different line styles and colors in panels (a) and (b) indicate the places where the four microeruptions are from. Each number corresponds to the number given to each microeruption in this paper.}
\label{fig:fig1}
\end{figure}

\begin{figure}
\epsscale{0.8}
\plotone{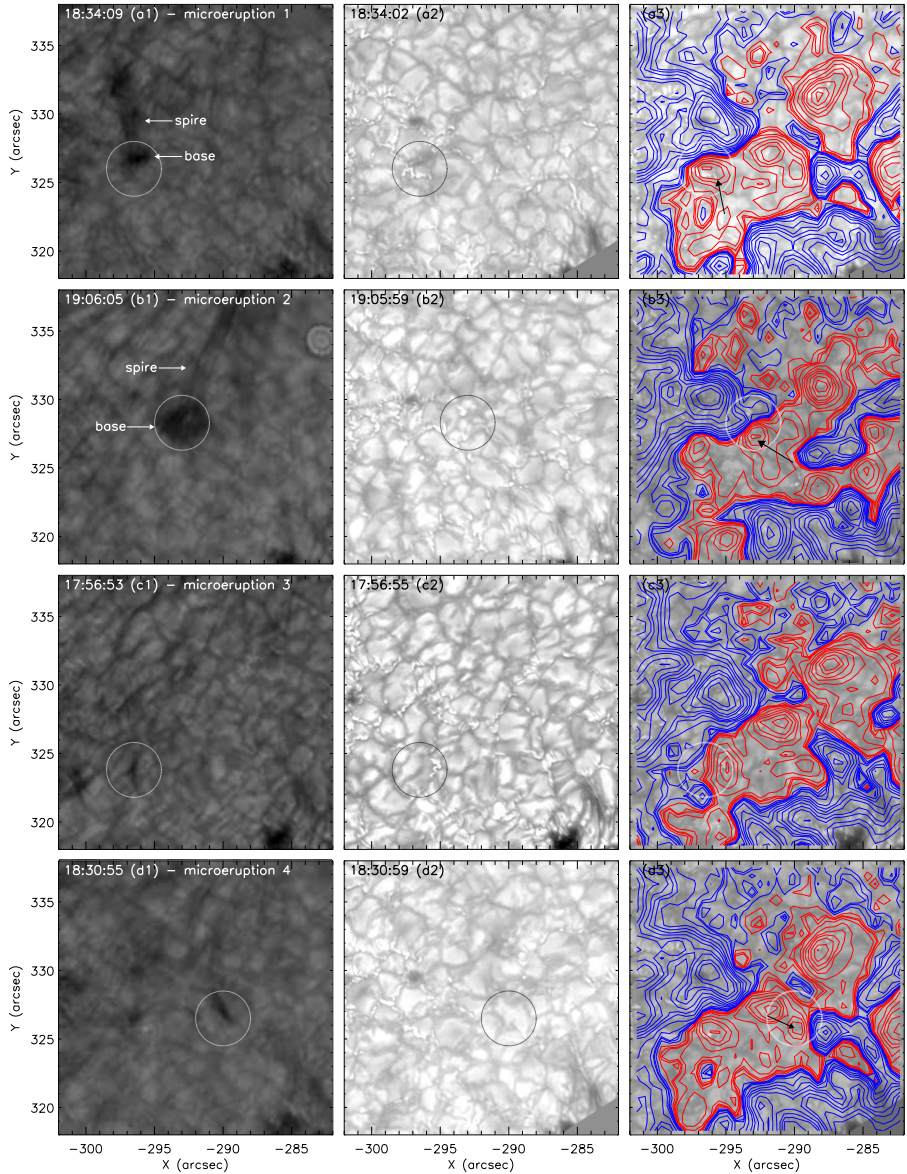}
\caption{The left column gives 4 separate snapshots for the four microeruptions (with two microjets in panels a1 and b1), as observed with He \text{\small{I}} 10830 \AA\ narrowband filtergrams.  The middle and right columns are corresponding broadband filtergrams at TiO 7057 \AA\ lines and the same filtergrams overlaid with contours of LOS magnetic field. The red and blue contours in the magnetograms represent positive and negative LOS magnetic field with the levels of  $\pm$ 7, $\pm$ 9, $\pm$ 15, $\pm$ 30, $\pm$ 50, $\pm$ 70, $\pm$ 100, $\pm$ 150, $\pm$ 200, $\pm$ 300, $\pm$ 400 G, respectively. The circles in each row have the same position on the maps and are used to relate the 4  microeruptions with the context of surrounding granules and magnetic field on the photosphere. The two arrows in panels (a3) and (b3) point to the moving magnetic elements being associated with microjets 1 and 2. The arrow in panel (d3) points to the emerging magnetic element being associated with microeruption 4. An online animation of the figure is available, it lasts for 103 minutes from $\sim$ 17:35 UT to 19:18 UT.}
\label{fig:fig2}
\end{figure}

\begin{figure}
\epsscale{1}
\plotone{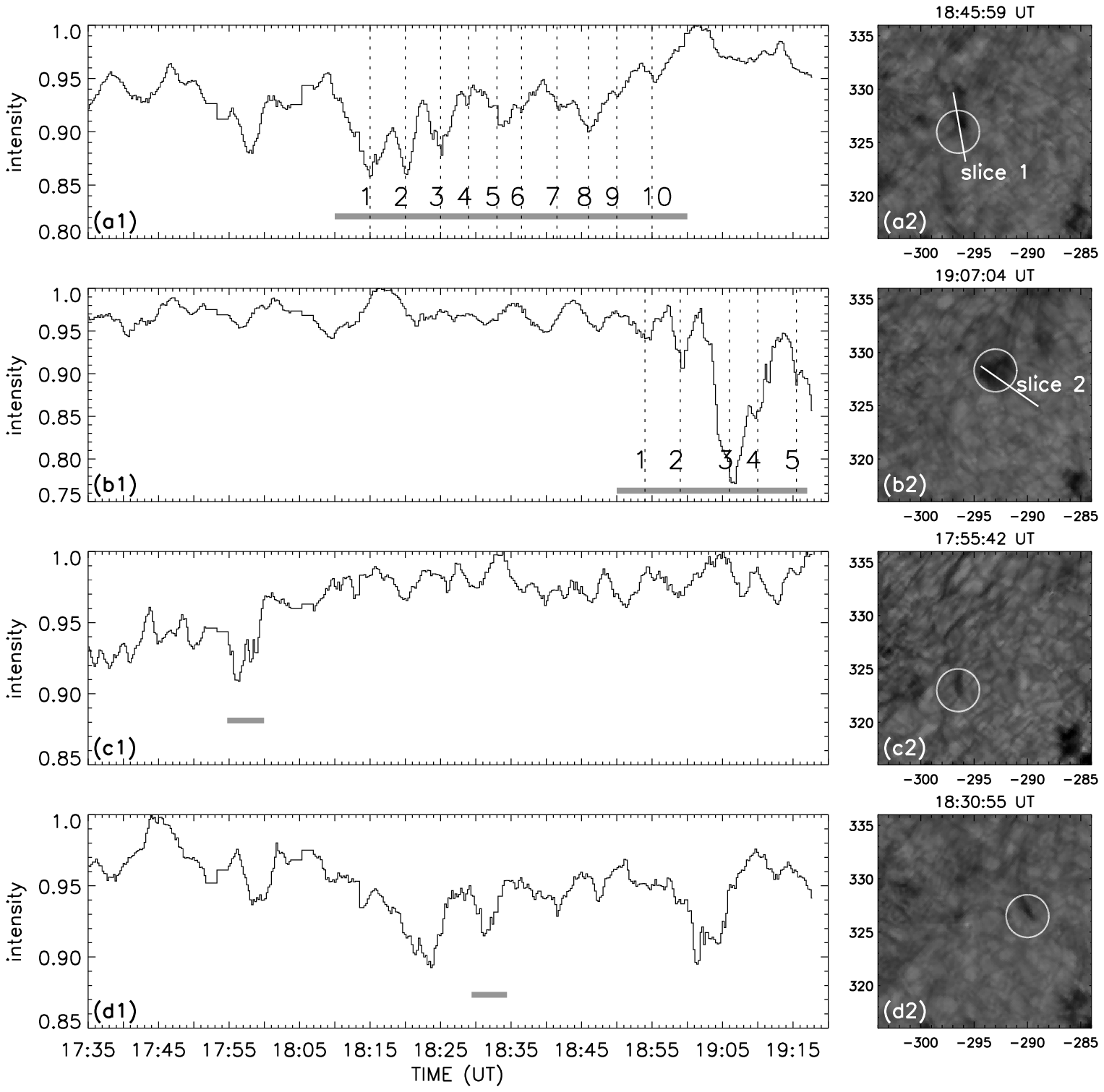}
\caption{The time profiles (normalized) for He \text{\small{I}} 10830 \AA\ brightness integrated from the circled areas given on the right column. The gray horizontal lines show the durations of the 4 microeruptions. Dashed lines in panels (a1) and (b2) give quasi-periodic enhancement of He \text{\small{I}} 10830 \AA\ absorption during the eruption processes of microjets 1 and 2. Slices 1 and 2  indicate the positions used to make the time distance diagrams in Figs. 5 and 7. An online animation of the figure is available, it lasts for 103 minutes from $\sim$ 17:35 UT to 19:18 UT.}
\label{fig:fig3}
\end{figure}

Figure \ref{fig:fig1} (a) gives an overview of the active region NOAA 11259 as observed by AIA 171 \AA, with the field of view (FOV) of $130{^\prime}{^\prime}$ $\times$ $130{^\prime}{^\prime}$. The black box corresponds to the area of the panel (b) with a FOV of $40{^\prime}{^\prime}$ $\times$ $40{^\prime}{^\prime}$, where a sample He \text{\small{I}} 10830 \AA\ image is given. The small boxes of different colors and numbers in panel (b) give the locations of the four microeruptions. It is worthy of noting that a moss region is on the left side of the picture, where many periodic small-scale mass ejections have been studied and they are reported to be powered by magnetoacoustic oscillations \citep{Hashim2021, Ji2021}. Fig. 2 along with its online animation gives detailed information of the four microeruptions on the He \text{\small{I}}  10830 \AA\ images and their positions (circled area) in the context of granules and LOS magnetograms on the photosphere.  The images in the second column show that the locations of the microeruptions are subject to constant advection of granules and have no special appearance when compared with other places. However, the third column makes it apparent that all the microeruptions are located around magnetic polarity inversion lines (PILs). Furthermore, with the help of animation for Fig. 2, we can see that all microeruptions squirt out their material from the area of inter-granular lanes. Spires and bases in microeruptions 1 and 2 are occasionally visible (as shown in panels a1 and b1), so the two eruptions can be taken as microjets. \\

The temporal behaviors for the He \text{\small{I}} 10830 \AA\ intensity in the circled areas of the  two microjets and two microeruptions are given on the left column of Fig. 3. On the right column, four snapshots for the two microjets and two microeruptions are given. In this way, an online animation for Fig. 3 will enable us to carry out a detailed comparison between their temporal variation and spatial evolution. The gray horizontal lines indicate the duration of the 4 microeruptions. They can be classified into two categories: long-duration microjets and short term microeruptions. The two long duration microjets lasted $\sim$ 50 and 27 minutes respectively. During their existence, they burst periodically for a period of $\sim$ 5 minutes, as can be seen from the oscillations imposed on the light curves.  Certainly, the oscillations of the light curves will have the component of the p-mode  oscillations of 5 minutes on the photosphere, since He \text{\small{I}} 10830 \AA\ images contain information from the underlying photosphere. However, when we visually check the movie made for Fig. 3, we can see that each burst in the circled area is nicely associated with each valley period on the light curves.  In this sense, microeruptions 3 and 4 appear to be singular and last only one period ($\sim$ 5 minutes). The light curves for microeruptions 3 and 4 reflect the variation of He \text{\small{I}}  10830 \AA\ absorption modulated by the p-mode oscillations of 5 minutes on the photosphere. \\

We check the spatio-temporal evolution of magnetograms with the animation of Fig. 2, we see that the configurations of the PIL associated with microjets 1 and 2 are more stable. It is long and exists there throughout the entire period of the events. There are two moving magnetic elements  continuously push and squeeze the PIL northward (pointed with arrows in panels a3 and b3). The pushing movement continuously squeezes and distorts the adjoining negative magnetic elements.   \\

\begin{deluxetable*}{ccccc}
\tablenum{1}
\tablecaption{Parameters of microeruptions\label{tab:tab1}}
\tablewidth{0pt}
\tablehead{
\colhead{Microeruptions} & \colhead{Length$\times$Width} &
\colhead{Start time} & \colhead{End time} & \colhead{Duration} \\
\colhead{}   & \colhead{(arcsec)} &
\colhead{UT} & \colhead{UT} & \colhead{(min)}
}
\startdata
1   & $\sim$$3 \times 1$  & $\sim$18:10:00 &  $\sim$19:00:00& $\sim$50 \\
2  & $\sim$$3\times3$  & $\sim$18:50:00 & $\sim$19:17:00& $\sim$27  \\
3  & $\sim$$3 \times 0.5$ & $\sim$17:54:45 & $\sim$17:59:56&$\sim$5  \\
4  & $\sim$$3 \times 0.5$  & $\sim$18:29:25 & $\sim$18:34:25& $\sim$5\\
\enddata
\end{deluxetable*}

\begin{figure}
\epsscale{1}
\plotone{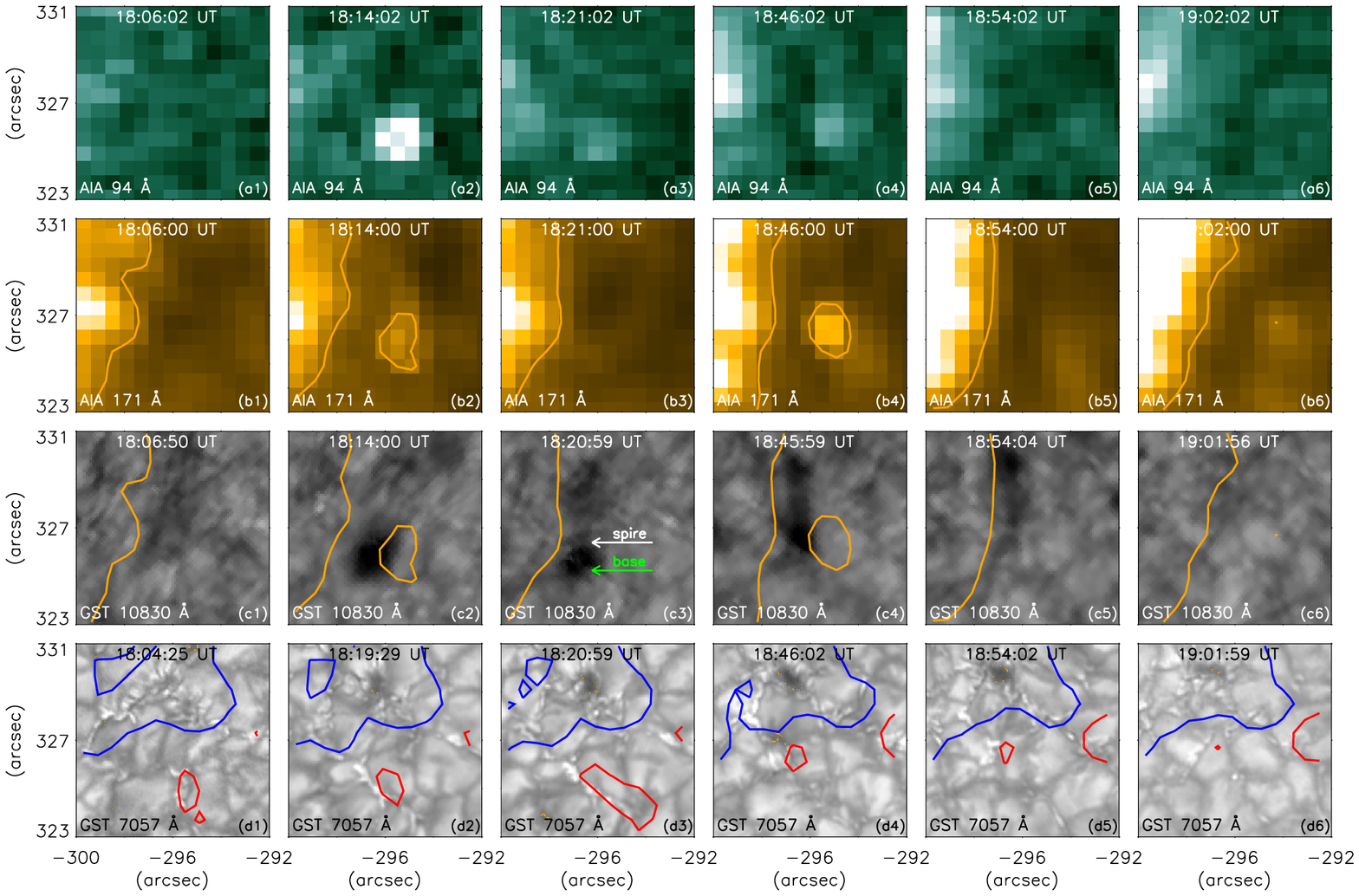}
\caption{Panels (a1-a6) and (b1-b6) in the top two rows give the observations for microjet 1 made by AIA at 94 \AA\ and 171 \AA.  Panels (c1-c6) in the third row show the repeated eruptions of microjet 1 as observed with He \text{\small{I}}  10830 \AA\ filtergrams, overlaid with simultaneous 171 \AA\ contours which have the value of 285 DN pixel$^{-1}$ s$^{-1}$.   The bottom row gives the underlying photosphere as observed with 7057 \AA\ filtergrams. The overlaid red and blue contours in the bottom row represent the LOS magnetic field with the values of +70 G and -70 G. White and green arrows in panel (c3) point out the spire and base of the microjet, respectively.}
\label{fig:surge1}
\end{figure}
\begin{figure}
\epsscale{1}
\plotone{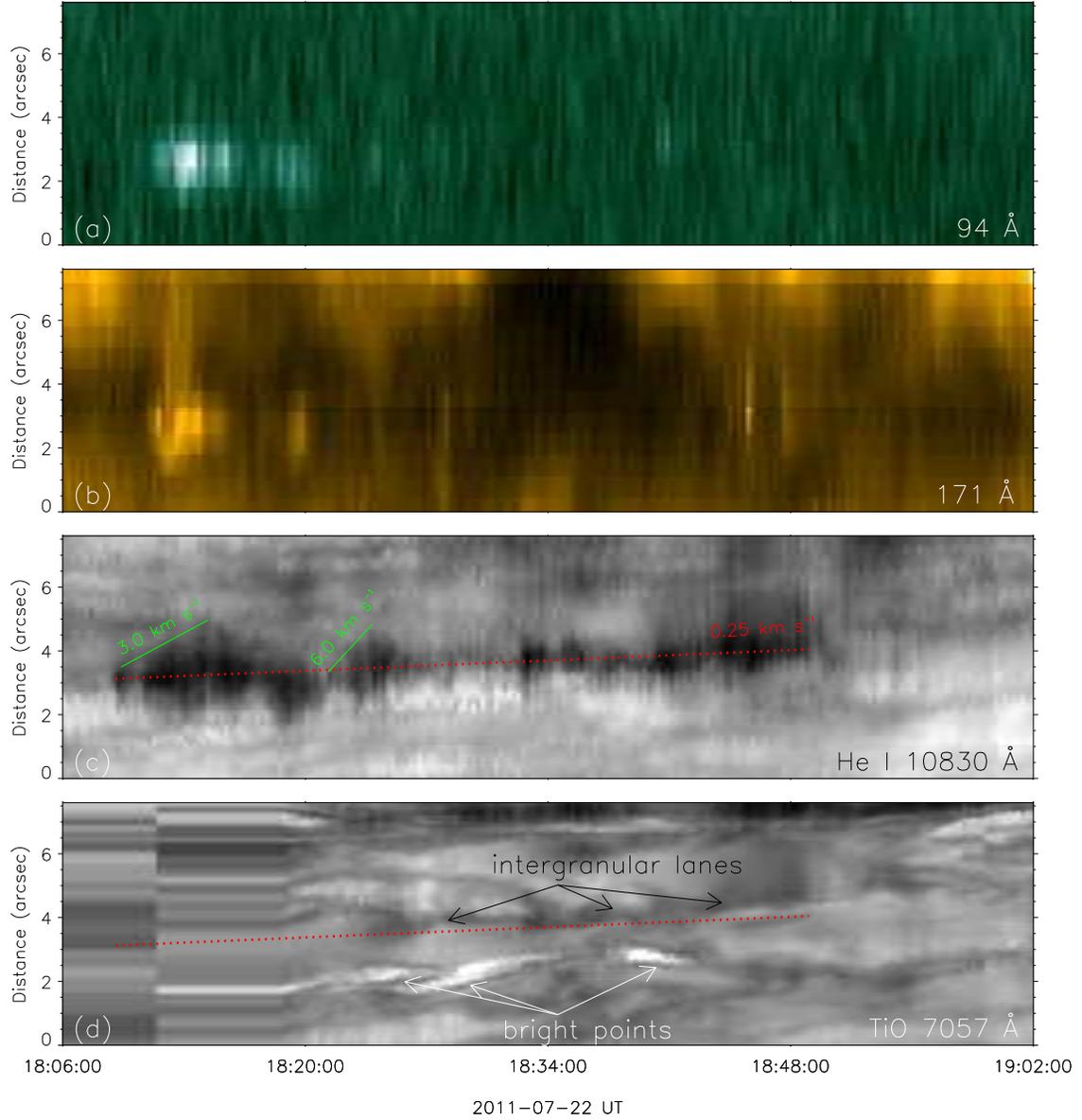}
\caption{From top to bottom row are the time-distance diagrams at 94 \AA, 171 \AA, 10830 \AA\ and 7057 \AA\ for microjet 1, which are obtained by stacking intensity images of slice 1 (Fig. 3) of corresponding wavelengths.  The two green lines in the third panel give fittings for the speeds of microjet 1 (3.0 km s$^{-1}$ and 6.0 km s$^{-1}$, respectively), while the red dotted line in the panel gives a fitting to the migration speed (0.25 km s$^{-1}$) for the base of the microjet. The same red dotted line is overplotted on the fourth panel, showing that the moving speed of the granules is equivalent to the base of the microjet. The dark arrows point out the intergranular lanes and the white arrows point to the bright points.}
\label{fig:surge1_slit}
\end{figure}

Seeking for their detailed dynamic behaviors associated with the magnetic field and granules, we will focus on the base of the two microjets. Fig. 4 and 6 give their temporal evolution as observed in AIA 94 \AA\ and 171 \AA, He \text{\small{I}} 10830 \AA\ and TiO 7057 \AA\ and LOS magnetic field. Microjet 1 has a full size of $\sim3{^\prime}{^\prime} \times 1{^\prime}{^\prime}$ with a base area of $\sim1{^\prime}{^\prime} \times 1{^\prime}{^\prime}$.  The He \text{\small{I}} 10830 \AA\ filtergrams in panels c1 to c6 in Fig. 4 again show that the microjet is recurrent from its initial generating to final fading. Combining with the TiO 7057 \AA\ photosphere images  and the contours of the magnetic field (panels d1-d6), it can be found that the positive magnetic field concentration (red contours) is decreasing during the whole period of microjet 1. The magnetic concentration is mainly situated in the intergranular lanes and associated with the bright points,  which are the important signatures for the stronger magnetic field. Meanwhile, the positive magnetic field and the bright points in intergranular lanes move toward negative magnetic field (blue contours). The base of the microjet is also moving accordingly and produces several co-spatial EUV brightenings at multiple temperatures (panels a2, a4, b2, and b4). EUV brightenings are observed as recurrent at the same place. As the microjet gradually disappears, the magnetic field at its root gradually weakens, and the bright points in the lane areas also gradually fade away (see panels c6 and d6 of Fig. 4). \\

Through a slice cutting across microjet 1 (slice 1 in Fig. 3), we have made its time-distance diagram that is shown in Fig. 5.   The time spans for Fig. 5 is from 18:06:00 to 19:02:00 UT (We have no TiO 7057 \AA\ data from 18:04:25 to 18:19:29 UT). From the time-distance image of He \text{\small{I}} 10830 \AA, it can be seen that the microjet squirts its material continuously from $\sim$ 18:10 to 19:00 UT. The ejection is quite gentle with a speed of less than 10 km s$^{-1}$(from Fig. 5 green lines).  Its base moves in accordance with the advection of a corresponding granule. The granules push the intergranular lane move northward toward the site of negative polarity, as can be seen from the TiO 7057 \AA\ time-distance image. The movement speed of the granules is estimated as 0.25 km s$^{-1}$.  During this process, several patches of recurrent brightenings are shown in the time-distance images of AIA 94 \AA\ and 171 \AA. Thus, the generating process of the microjet is well associated with the convective granules which squeeze and cancel the magnetic field of opposite polarity. During this process, the concentration of magnetic field in the convective granules gradually decreases and the bright points gradually disappear. \\

\begin{figure}
\epsscale{1}
\plotone{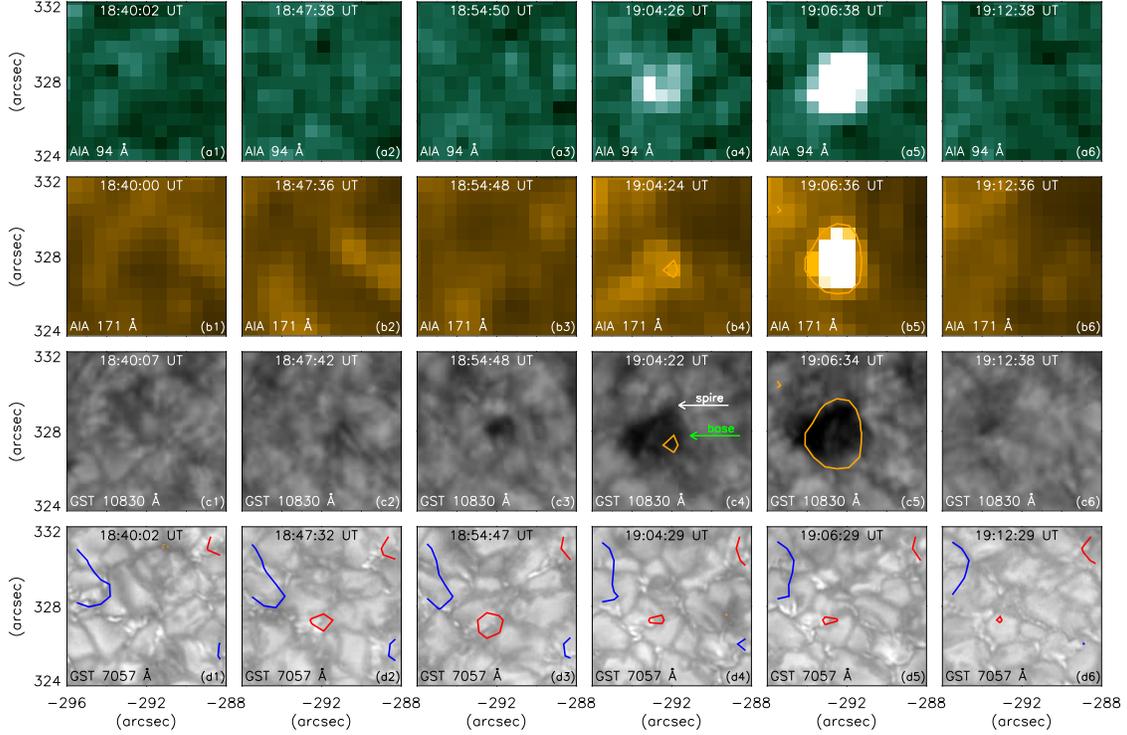}
\caption{Panels (a1-a6) and (b1-b6) in the top two rows give the observations for microjet 2 made by AIA at 94 \AA\ and 171 \AA.  Panels (c1-c6) in the third row show the repeated eruptions of microjet 2 as observed with He \text{\small{I}}  10830 \AA\ filtergrams, overlaid with simultaneous 171 \AA\ contours of the value of 285 DN pixel$^{-1}$ s$^{-1}$.  The bottom row gives the underlying photosphere as observed with 7057 \AA\ filtergrams. The overlaid red and blue contours in the bottom row represent the LOS magnetic field with the values of +150 G and -150 G. The spire and base of the microjet are pointed by white and green arrows in panel (c4), respectively.}
\label{fig:surge2}
\end{figure}

\begin{figure}
\epsscale{1}
\plotone{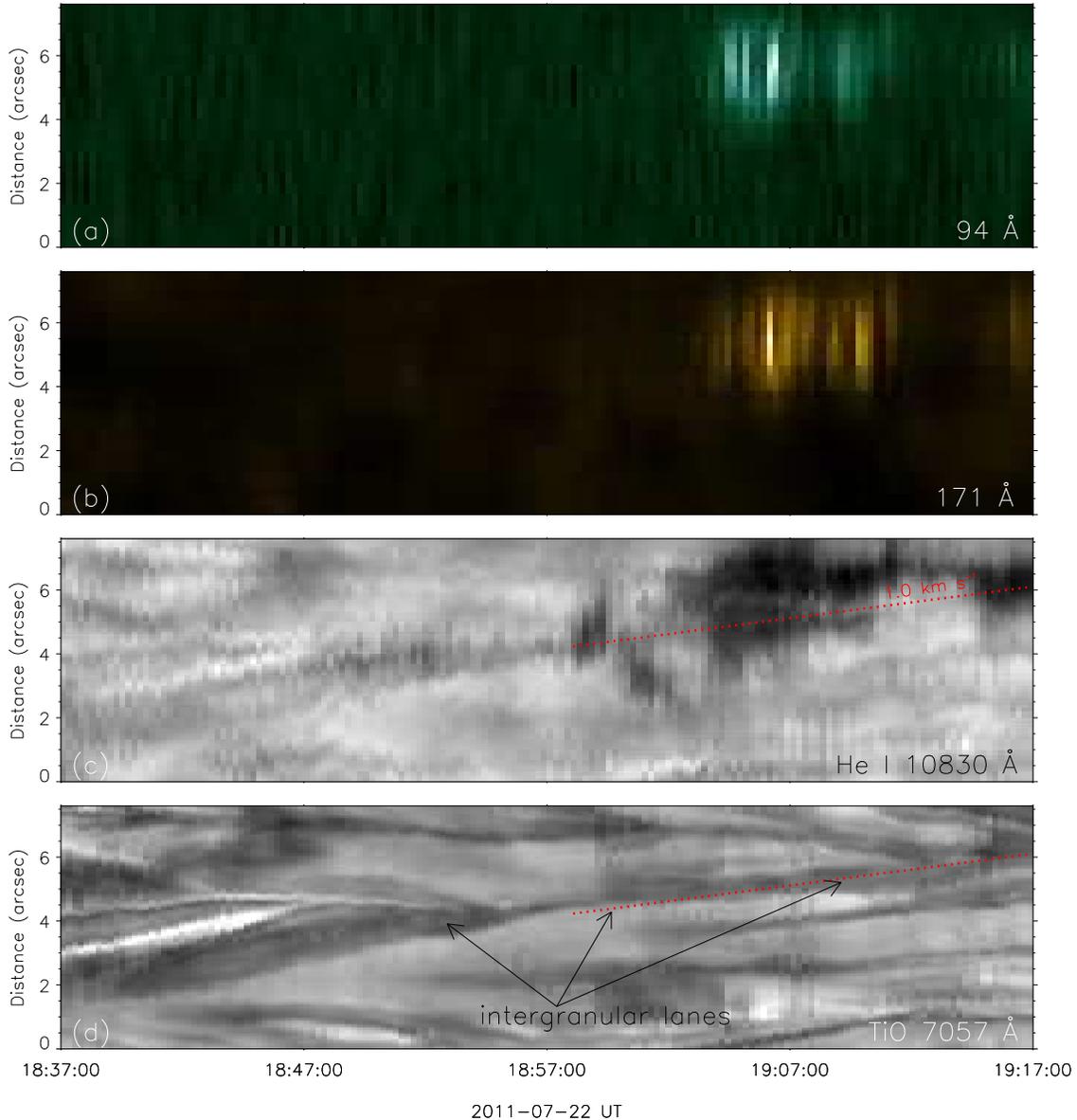}
\caption{From top to bottom row are the spacetime diagrams at 94 \AA, 171 \AA, 10830 \AA\ and 7057 \AA\ for microjet 2, which are obtained by stacking intensity images of slice 2 (Fig. 3) of corresponding wavelengths.  The red dotted line in the third panel gives a fitting to the migration speed (1.0 km s$^{-1}$) for the base of the microjet. The same red dotted line is overplotted on the fourth panel, showing that the moving speed is equivalent to the intergranular lanes which are pointed by dark arrows.}
\label{fig:surge2_slit}
\end{figure}

The second microjet has a base size of $\sim3{^\prime}{^\prime} \times 3{^\prime}{^\prime}$. We carried a similar analysis as for the first microjet, and the results are given in Figs. 6-7.  Figs. 6 and 7 show that the microjet is associated with the movement of the positive magnetic element. Co-spatial multi-temperature EUV brightenings at the base occur when the He \text{\small{I}} 10830 \AA\ absorption of the microjet get its maximum phase. Through the TiO time-distance image in Fig. 6, we again see that the movement of the magnetic field toward opposite polarity  coincides with the movement of the intergranular lanes. This behavior is highly similar to that of microjet 1, except that there are no obvious intragranular bright points.  \\

Microeruptions 3 and 4 are singular, lasting $\sim$ 5 minutes. They are more tiny with the width well below one arcsec  (Table 1). During their appearance, there are no detectable EUV brightenings. Also, there are no detectable magnetic field changes associated with microeruption 3. However, for microeruption 4, it is apparently associated with the emergence of a positive magnetic element (pointed by an arrow in panel d3 of Fig. 2). Furthermore, they are still rooted in intergranular lanes and magnetic PILs, being consistent with the corresponding results for microjets 1 and 2. 

\section{Discussions and Conclusions} \label{sec:con}

With high-resolution He \text{\small{I}} 10830 \AA\ narrowband images from GST,  we give an extensive analysis to 4 granule-sized microeruptions, two of which are identified as microjets. They all appear as enhanced He \text{\small{I}} 10830 \AA\ absorption.  The investigation of the microjets using high-resolution He \text{\small{I}}  10830 \AA\ narrowband images represents an extension for the current scale size spectrum of coronal jets down to the smallest spatial scales.  We have seen that high-resolution He \text{\small{I}} 10830 \AA\ imaging is an indispensable observation for various jet phenomena in the chromosphere. We can trace their roots directly to the underlying photosphere and its magnetic field. With the help from the He \text{\small{I}} 10830 \AA\ data and simultaneous high-resolution images of the photosphere, we report that these microeruptions are rooted in the area of intergranular lanes, which confirms previous results \citep{Ji2012, Zeng2013, Wang2021}.  Furthermore, they are situated on magnetic polarity inversion lines (PILs). \\

For the two microjets, we also find that they are caused by advection of granules which transport concentration of positive magnetic field to a relatively stable and long magnetic PIL. The moving positive magnetic field element persistently pushes and distorts the PIL, canceling the magnetic flux of the opposite magnetic field. In quiet regions and coronal holes, \cite{Panesar2016,Panesar2018,Panesar2020} and \cite{McGlasson2019} have found that magnetic flux cancellation at a PIL often makes a PIL-tracing minifilament and triggers its eruption. The minifilament eruption drives external reconnection of the minifilament's enveloping magnetic lobe with far-reaching magnetic field to produce a coronal jet. This phenomenon and model were first discovered and proposed by \cite{Sterling2015}. Meanwhile, microeruptions 3 and 4 were more like confined minifilament eruptions that didn't make jets. For a microeruption such as any of those in the present paper, a dark microeruption with or without a discernible microjet spire in He \text{\small{I}} 10830 \AA\ images, high-resolution H$\alpha$ or EUV images along with high-resolution magnetograms are needed to decide whether the microeruption is the product of a microfilament eruption. Aslo, many other works have been reported regarding how magnetic flux emergences and cancellations induce various recurring jets (e.g., \citealt{Chae1999, Zhang2014,Liu2016,Chifor2008,Zeng2013}, for a more comprehensive review, see \citealt{Raouafi2016}).  Therefore, our observation supports the magnetic reconnection scenario for the two microjets. \\

Coronal jets tend to be recurrent \citep{Joshi2017, Guo2013, Jiang2007, Cheung2015}. In most cases, they are not periodic. Nevertheless, \cite{Ning2004} and \cite{Doyle2006} reported a quasi-period of three to five minutes for repetitive transition region explosive events. \cite{Li2015} obtained the periods of 5 and 13 minutes from intensity variations at the base of a series of recurrent EUV jets.  With high-resolution narrowband imaging at He \text{\small{I}} 10830 \AA, \cite{Wang2021} reported quasi-periodic behavior of $\sim$ 5 minutes period for a small-scale jet. However, in the EUV passbands, the jet only appear as recurrent with no periodicity. Similarly, the long-duration microjets in He \text{\small{I}} 10830 \AA\ given in this paper are found to be quasi-periodic with the period of $\sim$ 5 minutes. The material ejections, which are supposed to be in cool chromosphere temperature, are quasi-periodic being occasionally accompanied by non-periodic yet recurrent multi-temperature brightenings in the EUV passbands. From Figs. 5 and 7, we can see that EUV brightenings are associated with the intenser cool material ejections, which may indicates a relatively stronger magnetic reconnection. Due to the constant advection of positive magnetic elements, the cool ejections are actually continuously modulated by the 5-minute oscillations, as can be seen from the two time-distance diagrams (Figs. 5 and 7). The observations strongly support a slow magnetic reconnection that is continuously driven by constant advection of granules and the slow magnetic reconnection is modulated by p-mode oscillation. \\

To explain the quasi-periodic behavior of repetitive transition region explosive events, \cite{Chen2006} carried out MHD simulations by imposing p-mode oscillations at the bottom boundary. The simulations successfully made the reconnection in the transition region proceed in periodically. However, for the microjets reported in this paper, magnetic reconnection is more likely to occur in the solar lower atmosphere.  Simulations of magnetic reconnection in solar lower atmosphere including the modulation of p-mode oscillations are needed. Especially, the magnetic reconnection should be able to produce multi-thermal components, which span a large temperature range from $10^4$  to $10^6$ K, like the simulation given by \cite{Ni2021}. Numerical simulations show that, for magnetic reconnection in the lower chromosphere, hot plasma with temperature more than $2.0 \times 10^{4}$ K  can be produced when plasma $\beta$ becomes relatively lower \citep{Ni2021, Cheng2021}. In this way, we can understand why the two singular microeruptions have no EUV counterparts. In the end, we also would like to conclude that, to understand the real nature of these He \text{\small{I}} 10830 \AA\ microjets, we need further investigations to reveal their relationship with mottles and dynamic fibrils, which have been extensively observed and studied. \\

\begin{acknowledgments}
We thank the anonymous referee for his/her helpful suggestions. SDO data were made available by the NASA/SDO AIA and HMI science teams. We thank the teams for providing the data. BBSO operation is supported by NJIT and US NSF AGS-1821294 grant. This work was  supported by NSFC grants 12003072, 11790302, 12073081 and 12173092.  The work of Wang is supported by Youth Fund of JiangSu No. BK20191108. This work is also supported by National Key R\&D Program of China 2021YFA1600500 (2021YFA1600502)
\end{acknowledgments}

\bibliography{sample631}{}
\bibliographystyle{aasjournal}

\end{document}